\documentclass[conf]{new-aiaa}
\usepackage[utf8]{inputenc}
\usepackage{cleveref}
\usepackage{graphicx}
\usepackage{float}
\usepackage{amsmath}
\usepackage[version=4]{mhchem}
\usepackage{siunitx}
\usepackage{longtable,tabularx}
\Crefname{figure}{Figure}{Figures}
\crefname{figure}{Fig.}{Figs.} 
\crefname{section}{Section}{Sections}
\crefname{subsection}{Section}{Sections}
\crefname{table}{Table}{Tables} 
\usepackage{multirow}
 \usepackage{color}
\usepackage{tabularray}
\usepackage{wrapfig}

\newcommand{\SELone}[1]{SEL\textsubscript{1} }
\newcommand{\SELtwo}[1]{SEL\textsubscript{2} }

\title{Low Thrust Electric Propulsion Mission Concepts For a 3-Meter Class Space Telescope}

\author{Yael M. Brynjegard-Bialik\footnote{Graduate Research Assistant, Department of Aerospace Engineering and Steward Observatory, AIAA Student Member} }
\affil{University of Arizona, Tucson, AZ-85719}
\author{Mohamed Nassif \footnote{Aerospace Systems Engineer III, AIAA Student Member}, Drew Latta\footnote{Aerospace Systems Engineer II }}
\affil{LSAS Tec, Vail, AZ-85641, USA}
\author{Neel Kunjur\footnote{Chief Technology Officer, AIAA Member}, Nicholas Rahaim\footnote{Manager of GNC, AIAA Member}, Paul DeTrempe\footnote{Senior GNC Engineer, AIAA Member}}
\affil{K2 Space, Torrance, CA-90501}
\author{Abhi Tripathi\footnote{Director of Mission Operations, UC Berkeley Space Sciences Laboratory} }
\affil{University of California, Berkeley, CA-94720}
\author{Isaac Saedi-Marghmaleki\footnote{Communications and Systems Engineer, Lunar and Planetary Laboratory}, Dillon O’ Reilly\footnote{R\&D Engineer/Scientist III, Astronomy and Steward Observatory}, Justin Hom\footnote{Postdoctoral Research Associate, Astronomy and Steward Observatory}, Patrick Ingraham\footnote{Associate Research Professor, Steward Observatory}, Jeffrey Kingsley\footnote{Associate Director, Steward Observatory, Department of Astronomy}, Buell T. Jannuzi\footnote{Director, Steward Observatory \& Department Head, Department of Astronomy}, S. Pete Worden\footnote{Laureate Professor University of Arizona, and  Chairman of the Breakthrough Prize Foundation. }, and Ewan S Douglas\footnote{Associate Professor, Department of Astronomy, AIAA Member}}
\affil{University of Arizona, Tucson, AZ-85719}

\begin{document}

\maketitle

\begin{abstract}
Space-based telescopes benefit from operating in stable orbital environments with reduced exposure to radiation and thermal fluctuations in order to minimize cost and maximize time for high-quality observations. Finding this ideal environment proves beneficial particularly for exoplanet discovery and characterization; direct imaging requires sub-nanometer wavefront stability and multi-hour observations, and transit detection requires parts-per-million photometric accuracy. Our team at University of Arizona's Steward Observatory and the Wyant College of Optical Sciences is evaluating various mission concepts for a 3-meter class telescope design, flying on a spacecraft bus equipped with a low thrust propulsion system. The presented mission analysis focuses on obtaining suitable transfer trajectories to the desired science orbit as well as understanding the radiation environment during the transfer, which is relevant for low thrust missions. The science analysis explores different operating orbits with the purpose of yielding maximum scientific return for detecting exoplanets. In this paper, we evaluate the use of a 2:1 lunar resonant orbit and a Sun-Earth L\textsubscript{2} halo orbit for our mission.
\end{abstract}

\section{Nomenclature}

{\renewcommand\arraystretch{1.0}
\noindent\begin{longtable*}{@{}l @{\quad=\quad} l@{}}

LRO  & Lunar Resonant Orbit \\
CR3BP  & Circular Restricted Three-Body Problem \\
SEL\textsubscript{1}/SEL\textsubscript{2} & Sun-Earth Lagrange Point 1/2 \\
GTO  & Geosynchronous Transfer Orbit \\
SSTO & Supersynchronous Transfer Orbit \\
STK  & Systems ToolKit \\
$ \Delta V$ & Delta-V \\
$I_{sp}$ & Specific Impulse (s) \\
$Re$  & Earth Radii \\
$i_{cr}$ &Critical Inclination for LRO \\
$AOP$ & Argument of Perigee \\
$AOP_{lp}$ & $AOP$ relative to the lunar plane \\
$SEM$  & Spacecraft-Earth-Moon angle \\

\end{longtable*}}

\section{Introduction}
Space telescope design relies on propulsion systems and orbit architectures that can balance mission cost, transfer time, operational flexibility, and scientific return, which all motivate our examination of low-thrust electric propulsion (EP) mission concepts for a large aperture, high-stability space telescope designed for general astrophysics and direct exoplanet detection. This mission concept leverages propulsion efficiency, optical stability, and science return to fully explore the design of an orbit trade study. \

For our investigation, we consider a notional 3-meter diameter space telescope. The telescope performance is strongly influenced by its operating environment; thermal fluctuations can degrade contrast and limit the ability to detect faint exoplanets near bright host stars. Because of this, it is important to operate in a thermally stable orbital environment and therefore essential to determine an optimal orbit for spacecraft configurations and mission design. \

Orbit selection involves balancing the stability requirements with communications/operations requirements. The Low Earth Orbit (LEO) environment presents challenges for operation of a space telescope; despite the advantageous proximity to Earth for communications and high data-rate downlink, the thermal environment is relatively unstable as evidenced by approximate 18 nm variations in the focus of the Hubble Space telescope over the course of an orbit \cite{lallo_experience_2012}. For this reason, we focus on more distant orbits without regular eclipses or significant Earth irradiance. Relative to LEO, the Sun-Earth Lagrange point orbits have improved thermal stability, which creates a more optimal environment for high-contrast imaging. Determining the best orbit distance to balance mission cost, communications availability, and thermal stability is essential for a successful space telescope mission. This trade was considered for the Nancy Grace Roman Space Telescope, which ultimately determined that the L2 environment provided the optimal stability and field-of-regard to complement its scientific goals \cite{demers_requirements_2015}. 

Pointing stability is another important consideration for astronomical instrumentation and communications. Due to larger moments of inertia, larger observatories tend to achieve improved pointing stability \cite{pong2018orbit}. Our mission has a relatively large telescope payload and pointing-sensitive science objective, meaning we are likely to require high pointing stability. This will mandate specific attention to mass balance and considerations for input disturbances, such as spacecraft jitter. 

Electric propulsion systems can use significantly less fuel mass than chemical propulsion systems for the same specific impulse, allowing for a larger budget for payload mass. Our proposed 3-meter telescope could be launched on commercially reusable rockets to an easily achievable launch drop-off orbit. This allows us to ensure the launch vehicle does not have to fly to a specialized orbit, lowering its probability of being recovered on the ground. Using an EP system allows us to fly to orbits that are used by current GEO missions. However, EP systems tend to require longer transfer times than chemical propulsion systems, which does factor into operations cost.

The data volume and downlink capability are other important factors that go into orbit selection. Astrophysics missions generate a great deal of data; for example, the Nancy Grace Roman Space Telescope is expected to downlink on the order of 11 Tbits/day of science data while JWST has a requirement to downlink 270 Gbits/day\cite{romanData,johns_james_2008}. As a result, the orbit can also impose more challenging communication requirements and increase costs. Closer orbits have higher data rates and are less dependent on complicated communication architectures. Ground stations are also more accessible from lower orbits, reducing mission cost. \

The University of Arizona has developed an unobstructed 3-meter diameter three-mirror anastigmat (TMA) space telescope concept, based on a spun-cast borosilicate primary mirror with precision thermal control \cite{west_space_2010,douglas_approaches_2023, nicolasSPIE2026} with an optical prescription evolved from previous TMA designs \cite{maier_design_2020,kim_large_2025} which will be detailed in forthcoming works.
An unobstructed design generally maximizes coronagraph performance \cite{mawet_vector_2010} and provides performance equivalent to a significantly larger conventional on-axis telescope \cite{philip_stahl_habitable_2020}.
The telescope concept fits within a Falcon-9 fairing and mass budget (Table \ref{tab:mass_budget}) and is also expected to be compatible with the New Glenn and upcoming Terran R launch vehicles.
This provides a baseline mass and telescope aperture which serve as the starting point for the orbit design and science yield analysis that will be presented here.

\definecolor{Gray}{rgb}{0.498,0.498,0.498}
\definecolor{Conifer}{rgb}{0.572,0.815,0.313}
\begin{table}
\centering
\caption{Observatory mass budget for a notional 3m observatory based on a UA 3m borosilicate mirror with a mass of approximately 900 kg.}
\label{tab:mass_budget}
\begin{tblr}{
  width = \linewidth,
  colspec = {Q[234]Q[233]Q[91]Q[387]},
  cell{1}{2} = {Gray},
  cell{1}{3} = {Gray},
  cell{1}{4} = {Gray},
  cell{11}{4} = {r},
  cell{12}{4} = {Conifer,r},
  vline{2-4} = {solid},
  hline{2,12} = solid,
}
 & Total CBE Mass (kg) & MGA & Predicted MEV Mass (kg) = CBE + MGA\\
 &  &  & \\
Telescope assembly & 2000 & 10\% & 2200\\
Solar array & 120 & 5\% & 126\\
Instrument assembly & 300 & 35\% & 405\\
Spacecraft & 1300 & 10\% & 1430\\
Propellent & 605 & 30\% & 787\\
Dry mass & 3720 & 12 \% & 4161\\
Wet mass & 4325 & - & 4948\\
 &  &  & \\
 & F9-Reusuable GTO &  & 5500\\
 &  & Margin: & 11\%
\end{tblr}
\end{table}

We explore three enabling technologies for this mission concept: electric propulsion, advances in mirror and wavefront control technology, and improvements in data transmission rates. These capabilities together support the viable plan for a significantly lower cost mission and allow for a quantitative study of the performance between two different orbit types. As a challenging and illustrative astrophysical problem, we will evaluate the  exoplanet detection rate of a 3-m telescope with a high-contrast coronagraph. Exoplanet yield modeling is an example of a science metric that enables the optimization of the mission, as many of the orbit engineering and design choices affect the science return to varying degrees. \

In this paper, we will apply the enabling technologies above to a comparative analysis of two candidate orbit configurations: a Sun-Earth L\textsubscript{2} halo orbit and a highly elliptical Earth orbit similar to that of the Transiting Exoplanet Survey Satellite (TESS). We first discuss the low-thrust trajectory design methods needed for the telescope to reach each orbit. We then outline the approach for estimating the exoplanet detection yield for the telescope in each orbit configuration. Then, we discuss the implementation of both mission architectures, taking into account propulsion requirements, transfer times, estimated costs, and estimated science return. We lastly make a comprehensive comparison between the two orbits, comparing mission cost and science return, and discuss implications for the future space telescope mission design. \

\section{Motivation}

\subsection{Trajectory design background}
The Sun-Earth Lagrange Point 2 has been a historically popular location to place space telescopes due to its unique characteristics of having the Earth, Moon, and Sun on one side, which ensures an uninterrupted view of the cosmos as well as a highly reduced thermal environment. A spacecraft placed in precise orbit around the \SELtwo{} point will continue to orbit this point with minimal orbit maintenance. For example, the JWST, Euclid, and GAIA missions have utilized this location by means of a halo orbit or a lissajous-type orbit around the L\textsubscript{2} point to perform scientific observations \cite{yu_launch_2014,euclid_collaboration_euclid_2025, perryman_gaia2005}.  Upcoming exoplanet science missions, such as the Roman Space Telescope and the PLATO and ARIEL missions, plan to operate in this location \cite{romanData, rauer_plato_2024,lueftinger_ariel_2023}. Our team will use the JWST science orbit properties as the operational orbit to design a transfer. We refer to this particular orbit in this paper as the \SELtwo{} halo orbit.

The orbit used by the Transiting Exoplanet Survey Satellite (TESS) mission is a 2:1 lunar resonant orbit (LRO). It was the first of its kind to utilize this type of orbit for an astrophysics mission. It is a highly eccentric and highly inclined orbit around Earth in which it completes two revolutions for every one lunar orbit. As the Explorer program had a smaller budget than a flagship mission, the mission team developed a unique orbit that can be reached with lower fuel requirements. The TESS mission had several science and mission requirements, including availability for wide field-of-view scans, that led it to choose this type of orbit for its mission. It has a perigee of 17 Earth Radii (Re) to reduce science downlink time with an apogee of 59 Re \cite{dichmann_trajectory_2014, dichmann_trajectory_2016}. The apogee is also placed away from the Earth-Moon plane to mitigate eclipses and lunar perturbations. Additionally, the TESS orbit employs a lunar resonance phasing condition where the Moon-Earth-Spacecraft angle oscillates around 90 degrees at apogee, which ensures minimal perturbations from the Moon and eliminates the need for any orbit maintenance while ensuring the trajectory remains above the GEO belt for 30 years \cite{dichmann_trajectory_2016}. Due to the long term stability characteristics exhibited by the IBEX mission in a 3:1 lunar resonant orbit, similar family orbits became favorable for the TESS mission \cite{carrico_ibex_nodate}. The TESS spacecraft was equipped with a chemical propulsion system and utilized a lunar gravity assist to transfer to its final orbit after launching on a Falcon 9. We will refer to the TESS-like orbit as the LRO for this paper.

\subsection{Electric Propulsion Background}
Electric propulsion (EP) is well-suited for missions that require a large total $\Delta V$ \cite{o2021electric} but cannot afford the propellant mass penalty of a purely chemical transfer. The core trade off is transfer time. EP replaces high thrust and short maneuver duration with low thrust, high specific impulse, and long-duration orbit raising. For transfers to energetically demanding destinations such as a TESS-like 2:1 lunar resonant orbit or a \SELtwo{} halo orbit, that trade becomes a mission-level design driver.

Hall-effect thrusters are a practical choice within this regime because they provide a useful balance between thrust, efficiency, and system complexity. Rather than treating the transfer as a sequence of impulsive burns, the spacecraft is modeled as continuously adding orbital energy over many revolutions \cite{kwon2016study}. As a result, the transfer time depends directly on the available thrust-to-mass ratio throughout the maneuver, not simply on the total required $\Delta v$. 

This coupling matters because EP performance is not constant across throttle settings. Thrust, specific impulse, and efficiency all vary with input power, and those variations directly feed into both transfer duration and propellant consumption, as well as propellant mass. This is particularly important for Krypton based Hall-effect thrusters, where performance can change significantly across the operating range. For that reason, the analysis in this work uses measured propulsion data as a function of input power rather than assuming a single fixed $I_{sp}$ or efficiency. This provides a more realistic estimate of the transfer-time and propellant-use trade space for the candidate science orbits.

\subsection{Radiation}
In order to minimize mission development time and cost, it is desirable to maximize use of commercial off-the-shelf (COTS) electronics. Generally, most COTS components tolerate a  total ionizing dose less than 10 krad \cite{sinclair_radiation_2013,hodson_recommendations_2020}. 
This motivates a close look at the time spent during transfer in the radiation belts. 

\subsection{Orbit impact on exoplanet detection}
Astrophysical science return depends on many factors. For this study we chose an exoplanet direct imaging survey example, as direct imaging of extrasolar systems at visible wavelengths is a top science priority \cite{decadal_pathways_2021}.
We want to determine how the selected orbit impacts the science return of an exoplanet direct imaging survey. Advances in telescope design, wavefront control, and coronagraph performance have all worked to improve exoplanet detection \cite{stark2014maximizing}. We wish to assess orbit architecture as a possible driver of exoplanet yield as well. Orbit geometry takes into account many observational constraints including target availability and environmental stability which are factors that directly impact the number of detectable exoplanets. \

Many different exoplanet detecting missions have chosen different orbital strategies, reflecting different priorities in cost, operational complexity, and science objectives. Proximity and propulsion requirements are important to take into account when choosing an orbit to reduce overall mission cost for mass and communications. We want to additionally factor in exoplanet detection yield as a function of the orbit chosen in order to gain new knowledge regarding orbital choice. \

As discussed, we focus on two orbit configurations in this work: \SELtwo{} and LRO. We want to assess their relative performance for a high-contrast, space-based exoplanet detection mission. Geometric observing constraints such as solar and lunar keepout angles and the resulting field of regard are key factors that influence detection capability \cite{spohn2024habitable}. Keepout angles define regions of the sky that must be avoided to prevent obstruction of the telescope by bright sources such as the Sun, Earth, and Moon. The field of regard represents the part of the sky accessible to the telescope at any given time. These constraints are connected to the orbit geometry to determine the observability of target stars. Moreover, the position of the telescope instantaneously in its orbit can impact the opportunities for detection. Viewing geometry is variable throughout the orbit, affecting target revisit rates, the window for integration time, and the ability to detect reflected light exoplanets at favorable phase angles \cite{stark2014maximizing}. Therefore, differences in the spacecraft position at any given time as a result of orbit choice could result in differences between the achieved exoplanet yield. \

Taking all of these considerations into account, we want to quantify how orbital selection and observing constraints affect exoplanet detection performance. By identifying which orbit maximizes target accessibility, we aim to inform future mission design choices and aid in optimizing scientific return for future space telescopes.

\section{Methodology}
\subsection{Trajectory Design Methodology}
In this section, we prescribe the methodology used to construct the transfer trajectory for both science orbits. For the 2:1 LRO case, we design the trajectory in a traditional way i.e. a forward transfer where we start from a launch orbit and end in final orbit insertion. However, for the \SELtwo{} halo orbit case, we utilize a backward transfer approach, a technique commonly used for orbits modeled under the circular restricted three body problem (CR3BP). Here we make use of weak stability boundary theory or dynamical systems theory to find appropriate transfers to the \SELtwo{} halo orbit from Earth. 
\subsubsection{Assumptions}
The following assumptions were used for both transfer trajectories.

\begin{itemize}
    \item A notional launch into Super Synchronous Transfer Orbit (SSTO);  in Mar 2028 
    \item Launch drop off parameters are typically 300 km x 60000 km x 28.5 deg
    \item Launch Wet Mass: 4500-5500 kg
    \item Nominal Dry Mass: 4000 kg
    \item Science Orbit operational lifetime : 5 years (minimum); 10 years (extended)
    \item 10-15 \% fuel margin available at science orbit insertion
    \item Propulsion System Parameters

    \begin{itemize}
        \item Hall Effect Thrusters
        \item Nominal Thrust: 1 N
        \item Specific Impulse($I_{sp}$): 2000 s
    \end{itemize}
\end{itemize}

\subsubsection{Integration of Electric Thrusters}

The electric propulsion (EP) methodology developed in this work was intended to provide a first order physics-based estimate of transfer time and propellant expenditure for a large spacecraft transferring from GTO to candidate science orbits. The approach was built around the central characteristic of EP mission design: the transfer is governed by continuous low-thrust acceleration, not impulsive maneuvering, so orbit growth, propellant depletion, and thruster operating point must be solved in a coupled manner.

The spacecraft was modeled as a variable mass system subject to continuous tangential thrust. At each time step, the instantaneous acceleration was computed from the thrust-to-mass ratio, while the spacecraft mass was reduced according to the electric-propulsion mass flow relation,
\[
\dot{m}=\frac{T}{g_0 I_{sp}}.
\]
The semi-major axis evolution was then approximated using a low-thrust energy-raising formulation, allowing the transfer to be propagated forward in time until the target orbit was reached or the propellant budget was exhausted. This framework captures the dominant system level EP coupling for large spacecraft. As propellant is consumed, the mass decreases, the effective acceleration rises, and the transfer rate changes throughout the maneuver.

Experimentally measured krypton Hall thruster performance data were used directly as functions of input power. Thrust, specific impulse, and efficiency were allowed to vary across the operating range, and interpolated performance curves were used to assess both measured and extrapolated operating points. This is important for the design of large-spacecraft EP because the propulsion system cannot be accurately represented by a single fixed $I_{sp}$ or efficiency value. Power processing, thruster efficiency, and achievable thrust are all strongly coupled to the selected throttle point.

From a mission design perspective, this methodology emphasizes the quantities that matter most for large EP spacecraft; available electrical power, thrust level, propellant mass fraction, and transfer duration. In this sense, the method is novel not because it replaces high-fidelity trajectory optimization, but because it ties experimentally derived thruster behavior directly to spacecraft-level transfer performance. This enables rapid but physically meaningful assessment of whether a given EP operating envelope can close the transfer to a demanding science orbit within acceptable time and mass constraints.

Our team utilized these results to obtain initial guesses of the total time-of-flight and fuel required at launch, which were then fed as input to \textit{Astrogator}. Due to the gravity losses associated with low thrust burns, one cannot use Tsiolkovsky's rocket equation to calculate the fuel required based on a $\Delta v$ value.

\subsubsection{Science Orbit Modeling: 2:1 LRO} \label{sec:sci_orb_mod_lro}

The 2:1 LRO orbit used by TESS is a result of resonance from the circular restricted three-body problem (CR3BP) and is an application of the Kozai-Lidov mechanism, which describes long term behavior of a highly-eccentric, highly-inclined long term behavior as referenced in \cite{dichmann_trajectory_2014,dichmann_trajectory_2016,parker_transiting_2018}. The TESS mission design team found several key parameters that the spacecraft needed to meet to enter the correct orbit which we adapted. These parameters are as follows:
\begin{wrapfigure}[20]{r}{.46\textwidth}
	\centering
    \includegraphics[width=0.45\textwidth]{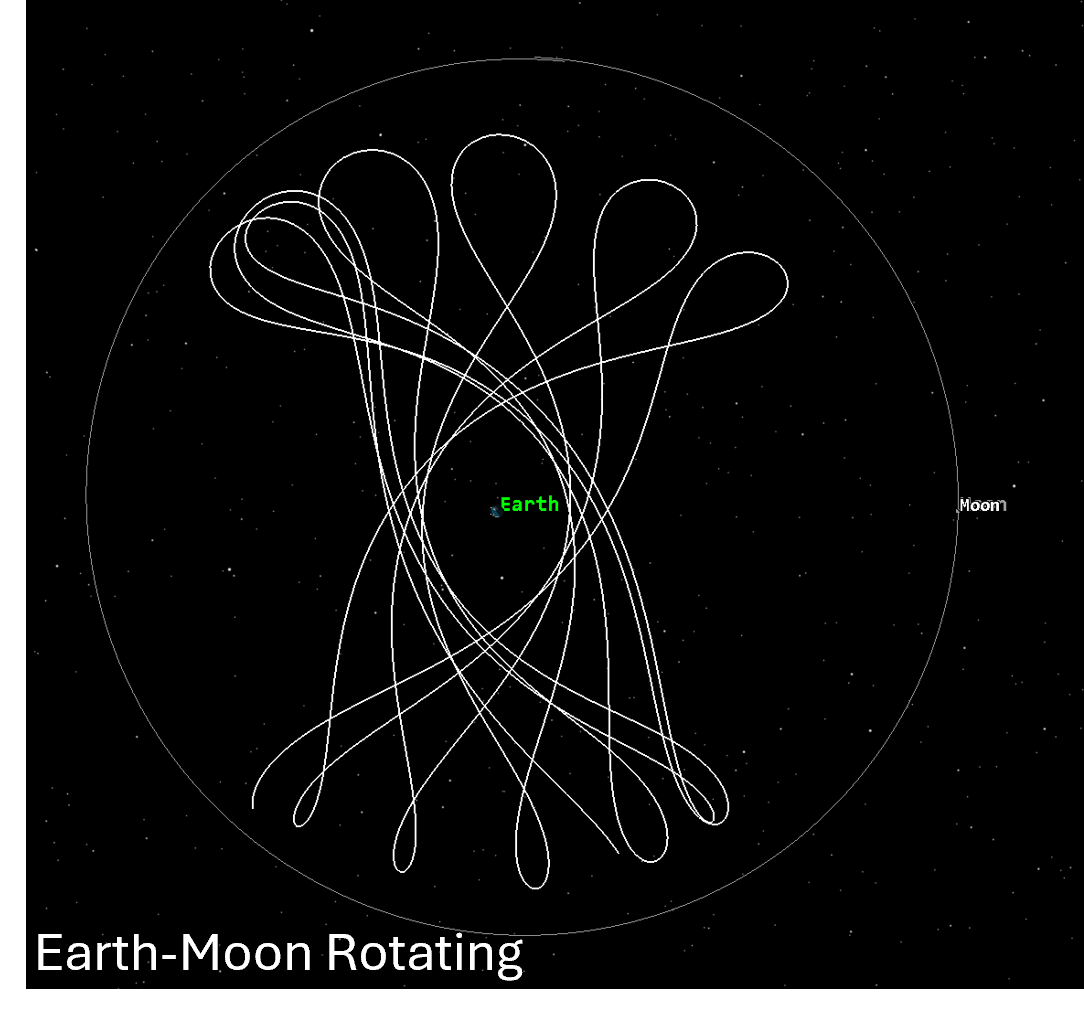}
    \caption{2:1 LRO propagated for one year (top view) showing the flower petal shape of the orbit in the Earth-Moon rotating frame}
    \label{fig:lro_1yr}
\end{wrapfigure} 

\begin{figure}[t!]
    \centering
    \includegraphics[width=0.975\linewidth]{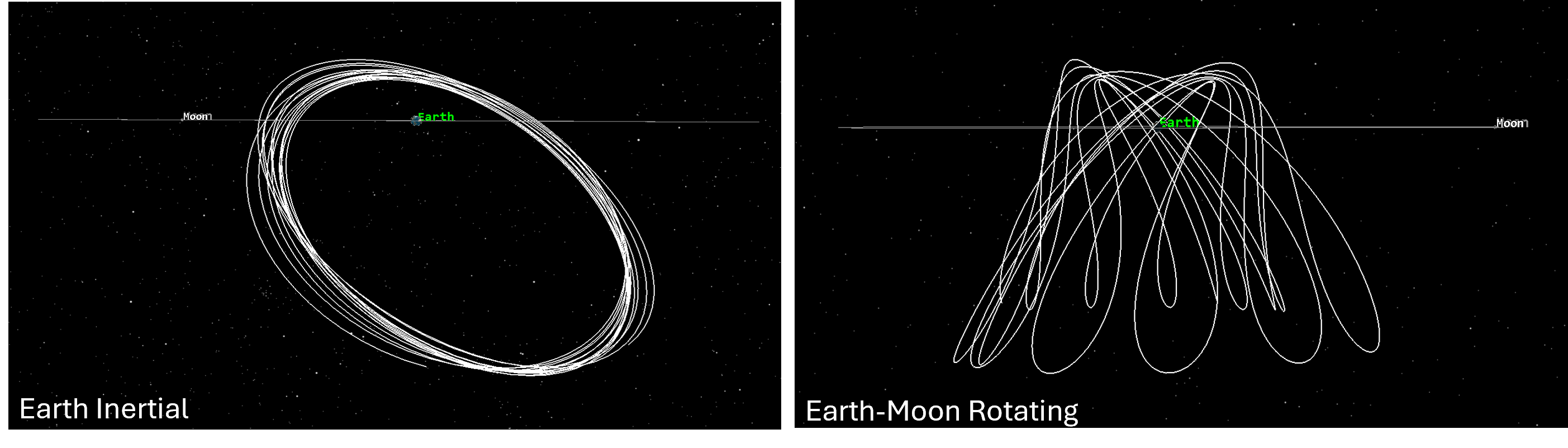}
    \caption{2:1 Lunar Resonant Orbit 3-D view in both Earth Inertial and Earth-Moon rotating frame showing the high inclination and out of plane components of the orbit}
    \label{fig:lro_1yr_3d}
\end{figure}

\begin{itemize}
    \item An initial perigee/apogee altitude of 108,800 km x 380,000 km
    \item Orbit inclined 37-40 deg from the Lunar Orbit Plane
    \begin{itemize}
        \item This was defined as the critical inclination $i_{cr}$, a parameter necessary for long-term resonance behavior
        \item The critical inclination is dependent on the orbit's inclination relative to the equatorial plane as well as the RAAN of the GTO/SSTO
        \item $i_{cr}$ is obtained by finding the angle between the lunar orbit plane normal and the spacecraft orbit plane normal. 
    \end{itemize}
    \item The argument of perigee relative to the lunar plane($AOP_{lp}$) must be close to 90 deg
      \begin{itemize}
        \item This ensures the apogee of the LRO is away from the Moon's orbit plane to minimize third body perturbations
        \item $AOP_{lp}$ is calculated as the angle from the intersection of the spacecraft orbit plane and the lunar orbit plane to the spacecraft orbit perigee. 
    \end{itemize}

    \item Ensure the Spacecraft-Earth-Moon ($SEM$ angle must be 0 $\pm$ 30 deg or 180 $\pm$ 30 deg at perigee and 90 $\pm$ at apogee.  
\end{itemize}
The 2:1 Lunar Resonant Orbit has been propagated for one year and is shown in \cref{fig:lro_1yr}. Additional views showing the out of plane components are shown in \cref{fig:lro_1yr_3d}.

\subsubsection{Science Orbit Modeling: \SELtwo{} Halo}

\begin{figure}[h!]
    \centering
    \includegraphics[width=0.85\linewidth]{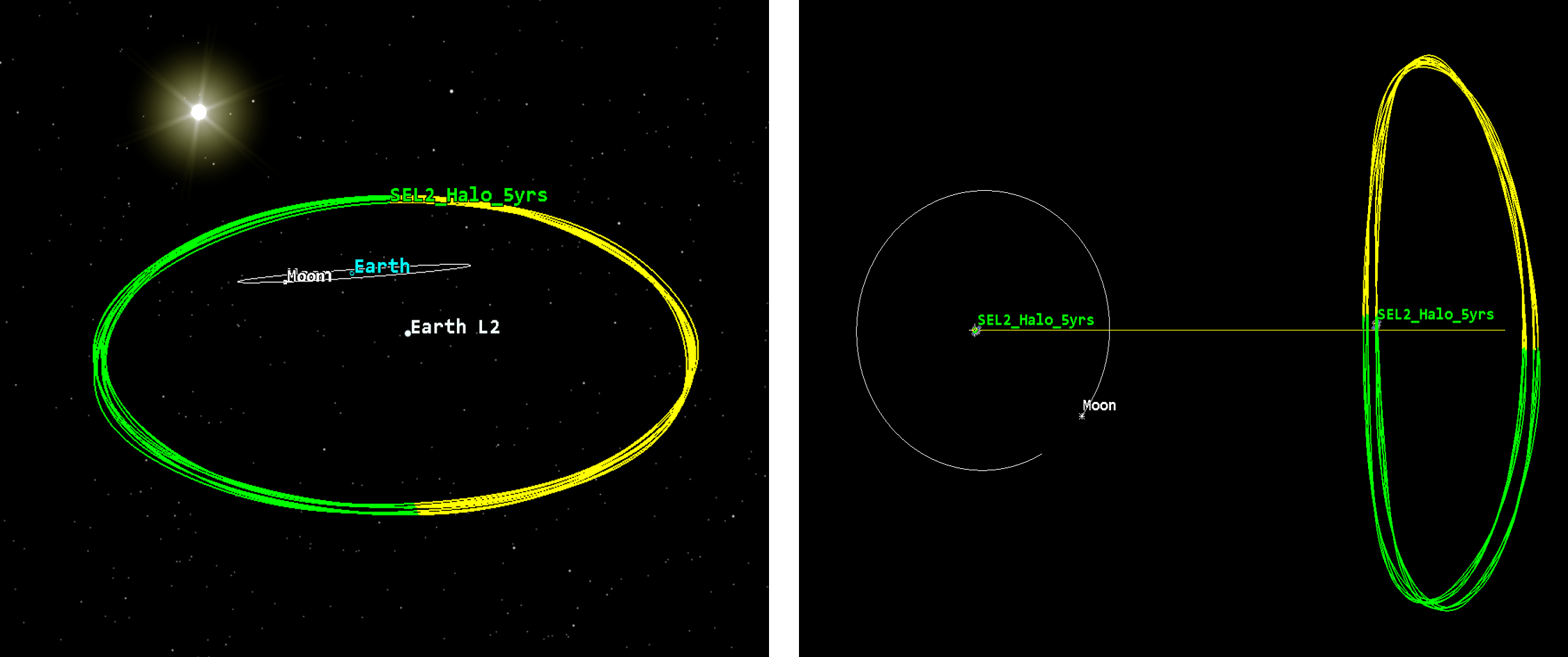}
    \caption{\SELtwo{} halo Orbit propagated for 5 years in the Earth-Sun rotating frame}
    \label{fig:sel2_fiveyears}
\end{figure}

The science orbit for this mission design was adopted from JWST's orbit design. JWST is in a Sun-Earth L\textsubscript{2} orbit in a halo/Lissajous orbit. For this study, we will make use of a \SELtwo{} Southern Halo Orbit. The orbit size was chosen to match JWST’s orbit size i.e. a max out of plane distance of 450,000 km and in-plane distance of 800,000 km found in \cite{yu_launch_2014,noauthor_james_2022}.
Initial conditions for these orbits are generated by finding solutions to the CR3BP problem in the Sun-Earth-Satellite case. JPL has published an online database that contains information on  a large number of initial states for Periodic Orbits in the CR3BP formulation. We used this database to find a Sun-Earth L\textsubscript{2} Southern Halo orbit  with the parameters mentioned above. The selected initial state belongs to orbit ID 455 in the database and is reflected in STK. Using \textit{Astrogator’s} differential correction tool, we correct the initial states in a full-force model to get a periodic halo orbit. The Receding Horizons method is used to correct the halo orbit for one full rev in the future. NASA's utilized this method to generate reference NRHO trajectories for the Gateway program \cite{williams_targeting_2017}. The five year differentially corrected halo orbit is shown in \cref{fig:sel2_fiveyears}. The orbit period is about 180 days i.e. the spacecraft completes one revolution of the halo orbit in 180 days. 

\subsubsection{Transfer Design Setup: Lunar Resonant Orbit}
For designing the transfer to a 2:1 Lunar Resonant Orbit, we break the mission into three phases. In Phase I, we target launch parameters to ensure the type of SSTO that can be launched into. In Phase II, we begin burns to spiral out in order to raise the spacecraft perigee above the outer radiation belts. This is done to minimize the time spent in the radiation belts. 
In the last phase, we adjust the perigee and apogee while targeting the required SEM angle at orbit insertion. 
A typical mission to GEO is as follows:
\begin{enumerate}
    \item Launch into an intermediate circular low earth parking orbit targeting a specific ascending node based on final required GEO longitude
    \item Coast until ascending node 
    \item Upper Stage (or kick stage) performs the burn to raise apogee and enter SSTO (this fixes the perigee at the ascending node and apogee at the descending node $\Rightarrow AOP \approx 0^{\circ}$)
    \item GEO spacecraft deployed and begins to circularize orbit and change orbit plane until GEO location reached
\end{enumerate}

In this particular case, as we are not concerned with transferring to a GEO, there is no requirement to target a specific ascending node nor do we have to do an apogee raising burn at the ascending node. In fact we would like to burn after passing the ascending node to ensure we have an argument of perigee $>0^{\circ}$. This allows us to ultimately target the final orbit plane $i_{cr}$ and $AOP_{lp}$ values described in \cref{sec:sci_orb_mod_lro} and reduce the fuel requirement by minimizing any plane change or orbit rotation maneuvers, both of which are fuel and time intensive. By carefully selecting the launch parameters and orbit parameters of the SSTO, we allow the spacecraft to focus only on changing the orbit size and not any other parameters.

The spacecraft is intended to be launched from Cape Canaveral. As the launch location and epoch determine the plane of the resulting parking orbit, i.e. the ascending node and orbit inclination, and the minimum inclination achievable by a launch from Florida is $\approx$ 28 deg, we vary the launch epoch to ensure the $i_{cr}$ of the SSTO is the range of allowable values\cite{dichmann_trajectory_2014,dichmann_trajectory_2016}. We use a Target sequence in \textit{Astrogator} to find the launch epoch to ensure our $i_{cr}$ is within 37-40$^{\circ}$. As we do not know which ascending node values can satisfy the $i_{cr}$ requirement, we instead simplify the targeting problem by letting STK identify opportunities for launch each day directly for $i_{cr}$. 

Once the satellite is in the intermediate circular low earth parking orbit, we need to find the right time to execute the apogee raising burn for the SSTO. The LRO requires an $AOP_{lp}$ value close to $90^{\circ}$ which itself is dependent on the ecliptic $AOP$. The location where the apogee raising burn occurs is what sets ecliptic $AOP$ of the resulting SSTO. We configure \textit{Astrogator} to find this burn location by using the time post ascending node as a control variable. This results in STK finding the appropriate coasting time in the low earth parking orbit before executing the SSTO burn, performed by the launch vehicle upper stage. 

As the spacecraft is deployed into an SSTO (300 km $\times$ 60000 km), we begin Phase II of the transfer– escaping the radiation belts. We thrust tangentially to raise our orbit until our perigee altitude is above 60,000 km. We are not concerned with the apogee or the shape of the orbit as doing so would increase the time spent traversing through the radiation belts. We also incorporate simplified eclipse conditions in this segment by adding stopping conditions in \textit{Astrogator} to stop thrusting and start coasting when exiting "Direct Sunlight" and resume thrusting when entering "Direct Sunlight". The final thrust duration required is calculated by \textit{Astrogator}. 

After the spacecraft converges on a solution for the above thrust duration, the resulting apogee has an altitude of $\approx 200,000$ km. We stop thrusting and then begin Phase III. The current orbit of the spacecraft is a stable inclined orbit with a perigee above GEO which mitigates GEO space traffic management concerns and the apogee is small enough that lunar perturbations are minimal allowing us to use this orbit as an intermediate parking orbit to coast in this orbit until we can identify a suitable moment to begin Phase III. This allows the onboard systems and engines to be re-calibrated and checked out after operating for $> 3$ months. We incorporate a mandatory coast sequence of a few orbit periods and then we identify the appropriate coast time needed to enter the final orbit by thrusting again. We seek to simultaneously raise the perigee and the apogee to the appropriate LRO size while also ensuring that the final SEM angle is within our required bounds. 

We do this by making use of technique in \textit{Astrogator} called "nested Target Sequences". An inner Target Sequence is used to raise the orbit to the appropriate orbit size while the outer Target Sequences work to identify how long to coast in the intermediate orbit before thrusting. To simplify thrust directions and have minimal thrust efficiency losses, we restrict the apogee raising burns to occur around the locations of the perigees ($-10^{\circ}<TA<10^{\circ}$) while the perigee raising burns occur around the apogees ($170^{\circ}<TA<180^{\circ}$). The exact start/stop locations are used as control parameters by the inner Targeter. The sequence loops multiple times until the inner Targeter identifies that the orbit size is close to what we need ($110,000 \times 380,000 $ km). After the inner Targeter has converged, the outer Targeter calculates the SEM angle at the end of the perigee/apogee raising burns at the apogee location and checks to see if it has satisfied the LRO requirements. It repeats the whole process with a new coast time until the SEM angle is within our bounds. 

The setup for calculating a transfer trajectory to the LRO has now been completed. The results are shown in the \cref{lab:lro_results}. 

\subsubsection{Transfer Design Setup: \SELtwo{} Halo}

\begin{wrapfigure}[19]{r}{.46\textwidth}
    \centering
    \includegraphics[width=0.78\linewidth]{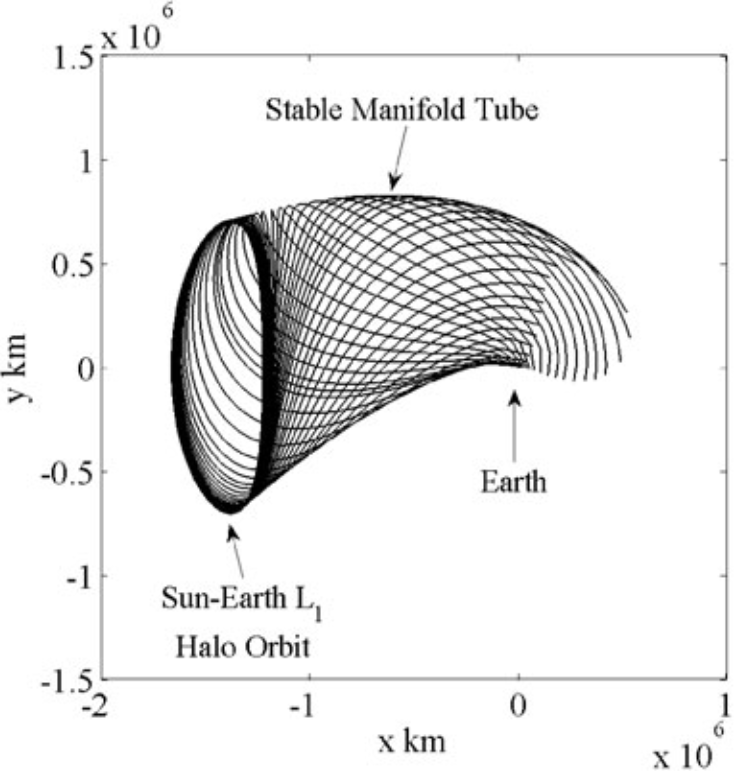}
    \caption{Stable manifolds of SEL1 halo orbit adapted from \cite{howell_representations_2006}}
    \label{fig:sel1_manifolds}
\end{wrapfigure}

Transfers to orbits that are a result of the CR3BP are typically evaluated using weak stability boundary theory/manifold theory/dynamical systems theory. More information on this can be found in \cite{parker_methodology_2014,parker_transfers_2014}. For a transfer to the \SELtwo{} halo orbit, we make use of a stable invariant manifold. This stable manifold is a ballistic pathway generated from the halo orbit backwards in time. The general idea is that if a satellite is put on this manifold at some time $T^-$, then at some time in the future at $T^0$, the satellite will enter the halo orbit ballistically. However, in an ideal world, there is some delta-V needed to place the satellite as well as to to enter the halo orbit, although the latter is generally minimal. This type of transfer is called a low-energy transfer– as in minimal fuel required– but it does tend to have long transfer times. An illustration of stable manifolds of the \SELone{} halo orbit taken from \cite{howell_representations_2006} is shown in \cref{fig:sel1_manifolds}.

All \SELone{} and \SELtwo{} missions have generally used this strategy to place satellites around L\textsubscript{1} or L\textsubscript{2}. We utilize this very strategy in this study to generate a transfer trajectory to a \SELtwo{} halo orbit. We will identify what the spiral out trajectory is needed to catch the manifold to enter the halo orbit. We do this by designing the trajectory backwards in time starting from a halo orbit through a trial and error method. In these types of backward shooting methods, it is typically a trial-and-error approach to visualize the solution space. Once we have an estimate of the manifold shape and direction we can then identify the targeting parameters to use in \textit{Astrogator} and have it converge to a solution. 
First, we coast backwards in the halo orbit, and perform a small burn in the direction of Earth. In the forward sense, this burn would be the halo insertion burn. Then we propagate backwards on the manifold and begin a thrust segment to ensure our trajectory goes back to Earth  and is captured into an Earth orbit with a perigee around 150,000 km. We also use this thrust segment to slightly adjust our inclination at Earth capture to be close to the GTO/SSTO inclination of 28.5 deg. Once we are captured into an Earth orbit, we begin to lower our orbit until our perigee is around 60,000 km. This mirrors our approach from the LRO transfer where we raise our initial low perigee altitude to 60,000 km to avoid the radiation belts. We let \textit{Astrogator} find the thrust directions and duration necessary to lower the vantage while ensuring that the inclination does not drift too much. 

After this step, we include a coast period before beginning to lower our orbit once again. Here we focus on lowering the parameter to the SSTO of around 300 km. Similar to the LRO transfer, we incorporate eclipse conditions during this orbit lowering segment. To ensure the SSTO we obtain after backward transfer is consistent with an actual SSTO that can be launched into including the correct ascending node and argument of perigee, we create another Satellite object in STK and design a forward trajectory segment that starts from the launch site and transfer in a forward propagation mode to an SSTO. We then have the backward propagating Satellite object target these parameters during its orbit lowering phase. 

A lunar flyby during the backward propagation was considered briefly, but targeting a lunar flyby would require additional calculations to identify what plane changes or orbit raise/lowering is required by the flyby to ensure the trajectory converges. In the interest of simplifying calculations, we chose not to consider a lunar flyby at this time. However, follow-on work on this project will include studying a lunar flyby to transfer to the \SELtwo{} halo orbit.

\subsubsection{Radiation work}
STK's Space Environment and Effects Toolkit (SEET) module was utilized for preliminary radiation analysis to predict the dose that each of the transfer trajectory studied would receive in the time before exiting the Van Allen radiation belts. For each trajectory, the accumulated radiation dose was computed for 5 mm and 10 mm aluminum shielding, testing both the NASA and Combined Release and Radiation Effects Satellite (CRRES) computational modes for 4 total configurations for each trajectory. The radiation analysis for both transfer orbits was done in a separate scenario. The converged transfer trajectory ephemeris was exported as *.e file and then imported into a separate scenario.

The SEET properties for the STK satellite objects were configured as follows:
\begin{itemize}
    \item \textbf{SEET Radiation settings:}
    \begin{itemize}
    \item \textbf{Computational Mode:} CRRES, NASA
    \item \textbf{Detector Type:} Silicon
    \item \textbf{Detector Geometry:} Semi-Infinite slab
    \item \textbf{Dose Integration Step:} 6 s
    \item \textbf{Dose report step:} 6 h
    \item \textbf{Shielding Thicknesses:} 5 mm, 10 mm
\end{itemize}
\item \textbf{SEET Environment Magnetic Field Model:}
\begin{itemize}
    \item \textbf{Main Field:} Fast IGRF
    \item \textbf{External Field:} Orson-Pfitzer
    \item \textbf{IGRF Update Rate:} 1 day
\end{itemize}
\end{itemize}
Properties that are not specified were left as the default settings.
The SEET analysis was computed for the predicted trajectory ephemeris for a time interval starting at the time of launch and ending  2 months after the spacecraft's perigee altitude is raised above 60,000 km.

\subsubsection{Communications/Coverage analysis at final orbit}
Our team developed a communications link-budget model to evaluate deep-space payload data downlink performance for multiple frequency/radio types for our space telescope mission. To ensure that our analysis was grounded in heritage systems with high TRL, we considered communications systems flown on JWST (Ka-band) and GAIA (X-band) and established baseline link budget models using publicly available information on those missions\cite{noauthor_jwst_nodate-1,johns_james_2008, perryman_gaia2005}. Both GAIA and JWST used S-band radios for tracking and telemetry which is the standard for deep-space missions and we intend to do the same.

In this study, we are only focused on evaluating the payload radio downlink performance. After validating the baseline models, we varied key mission parameters such as spacecraft-to-Earth distance, ground-station dish diameter, and spacecraft transmitter power to examine their effect on link performance. The model was also used to determine the maximum supportable data rate as a function of distance while keeping link margin constant at 3 dB, which is a common assumption for preliminary communications analysis. For that portion of the study, the ground station dish size was kept constant at 20 meters, since this was considered the most likely usable antenna size from commercially available ground stations. All other RF parameters (modulation scheme, forward-error correction algorithm, antenna, radio, and all other RF hardware) were kept the same as GAIA values for X-band cases and as JWST values for Ka-band cases. At this time our team has not yet identified the spacecraft antenna size as this depends on the frequency selected. As such, we refer to the onboard antenna sizes from GAIA and JWST for our study. The maximum data-rate analysis was performed first using the heritage transmitter powers from GAIA and JWST, and then repeated for a higher power 160 W transmitter case to assess the performance benefit of a possible higher-power RF subsystem due to higher power availability on our spacecraft bus. This approach provided a practical way to compare architecture options and understand how achievable science return changes with mission geometry and communications system assumptions. The trajectory team provided the station-spacecraft ranges as well as the available access times to various commercial ground stations. 

\subsection{EXOSIMS Methodology}
\subsubsection{Orbit implementation, chosen universe and target list}
Our exoplanet yield estimates are computed with the open-source exoplanet imaging mission simulator EXOSIMS version 3.6.5 \cite{SavranskyEXOSIMS}. Within EXOSIMS we constructed specifications for a simulated universe using a consistent star catalog, target list, planet population, and completeness model. Our specification inputs are shown in \cref{tab:exosims_specs}. We used the SAG13 simulated universe to simulate the exoplanet occurance rates and distributions \cite{SAG13}. Within our specifications, we also included the design parameters for the telescope system as the instrument specification in order to properly represent the instrument. We then modeled the telescope in both the \SELtwo{} halo orbit and the LRO separately. The \SELtwo{} halo orbit describes approximately a 6 month halo orbit which is then propagated for the duration of the mission, generated from a MATLAB state vector. This functionality is built in to EXOSIMS as the default Observatory. Using a similar method, we constructed an Observatory file for the TESS orbit drawn from an ephemeris produced by JPL Horizons \cite{Giorgini_Horizons} and converted it to the correct MATLAB orbit file. We propagated these orbits for five years (the approximate mission duration) and kept all other specifications constant for simulation. We assumed that the mission was completely dedicated to exoplanet detection in order to maximize yield. Important factors that will directly impact the exoplanet yield are the orbit geometry, continuous viewing zones, thermal stability, keepout angles, and the duty cycle, which are all orbit-dependent and handled through EXOSIMS.

\begin{table}[h!]
\centering
\caption{EXOSIMS Specifications (constant between orbit configurations)}
\begin{tabular}{|l|l|}
\hline
\textbf{Category} & \textbf{Input} \\ \hline
Simulated Universe & SAG13 Universe \\ \hline
Star Catalog & EXOCAT1 \\ \hline
Completeness & Brown Completeness \\ \hline
Mission Life & 5 years \\ \hline
Pupil Diameter & 3 meters \\ \hline
Contrast & $1 \times 10^{-10}$ \\ \hline
Solar Keepout Angles & $[45, 180]$ \\ \hline
Earth Keepout Angles & $[45, 180]$ \\ \hline
Maximum Allowed Target Visits & 5 \\ \hline
Magnitude Limit of Observed Stars & 10 \\ \hline
Number of Observed Stars & 34 \\ \hline
\end{tabular}
\label{tab:exosims_specs}
\end{table}

\subsubsection{Determining science yield and keepout}
Revisiting target stars is essential for finding exoplanets; if systems are incompatibly arranged at the first time of observation, it is important to revisit systems and ensure that planets were not missed. Our mission allows a maximum of five visits to each star. The yield calculation includes the number of planets detected using each Observatory file (one for each telescope orbit), represented by the mean and standard deviation of the data. 
We also construct keepout maps and completeness maps in order to demonstrate the obstruction of the telescope at any given time and the overall functionality of each orbit for the mission goals. The completeness represents the probability of detecting exoplanets at each target and is thus essential for determining this functionality. These maps together indicate target availability and observability for the mission duration and provide a valuable measure of orbit utility for the exoplanet detection mission to enable a comparison between the options.

\section{Overall results and comparison}
\subsection{Results of trajectory design, satisfying constraints}
\cref{tab:traj_results_summary} shows a summary of the trajectory design results for the two orbit cases. While the \SELtwo{} halo orbit has a higher fuel requirements and longer transfer times, the lower $\Delta V$ value comes from spending less time in Earth's gravity well. 50\% of the transfer duration for the \SELtwo{} halo orbit is attributed to its coasting phase where it is ballistically being captured into a halo orbit. 
\begin{table*}[htbp]
\centering
\caption{Summary of Trajectory design results }
\label{tab:traj_results_summary}
\begin{tabular}{|l|c|c|c|}
\hline
\textbf{Parameter} & \textbf{Units} & \textbf{LRO} & \textbf{\SELtwo{}} \\
\hline
Initial Orbit & -- & SSTO & SSTO \\
Transfer Time & days & 190 & 315 \\
Total $\Delta v$ & m/s & 1350 &  1170\\
Total thruster on-time & days & 138 & 152 \\
Fuel remaining at science orbit insertion & kg & 144 & 100 \\
Transfer fuel required & kg & 605 & 665 \\
\hline
\multicolumn{4}{|c|}{\textit{Orbit Characteristics}} \\
\hline
Orbital Period & days & 180 &  13\\
Station-Keeping $\Delta v$ (annual) & m/s/yr & 0 &  5\\
\hline
\end{tabular}
\end{table*}

\subsubsection{LRO Results} \label{lab:lro_results}
\begin{figure}[h!]
    \centering
    \includegraphics[width=0.95\linewidth]{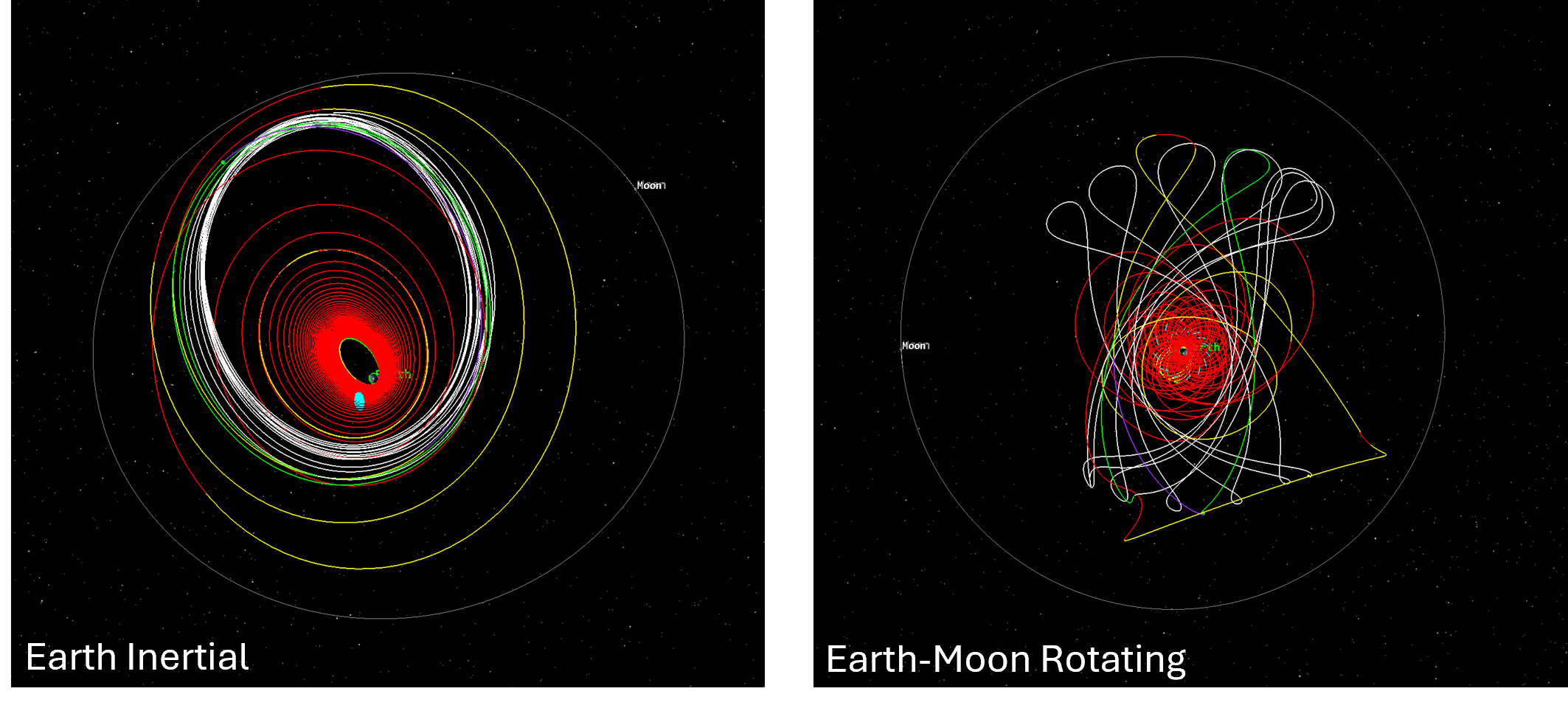}
    \caption{Converged transfer trajectory to a 2:1 LRO from SSTO in both Earth Inertial and Earth-Moon rotating frames. Thrusting segments are shown in red, and eclipse locations are shown in blue while other colors are coasting segments.}
    \label{fig:lro_trajectory_converged}
\end{figure}

\begin{figure}[t!]
    \centering
    \includegraphics[width=0.95\linewidth]{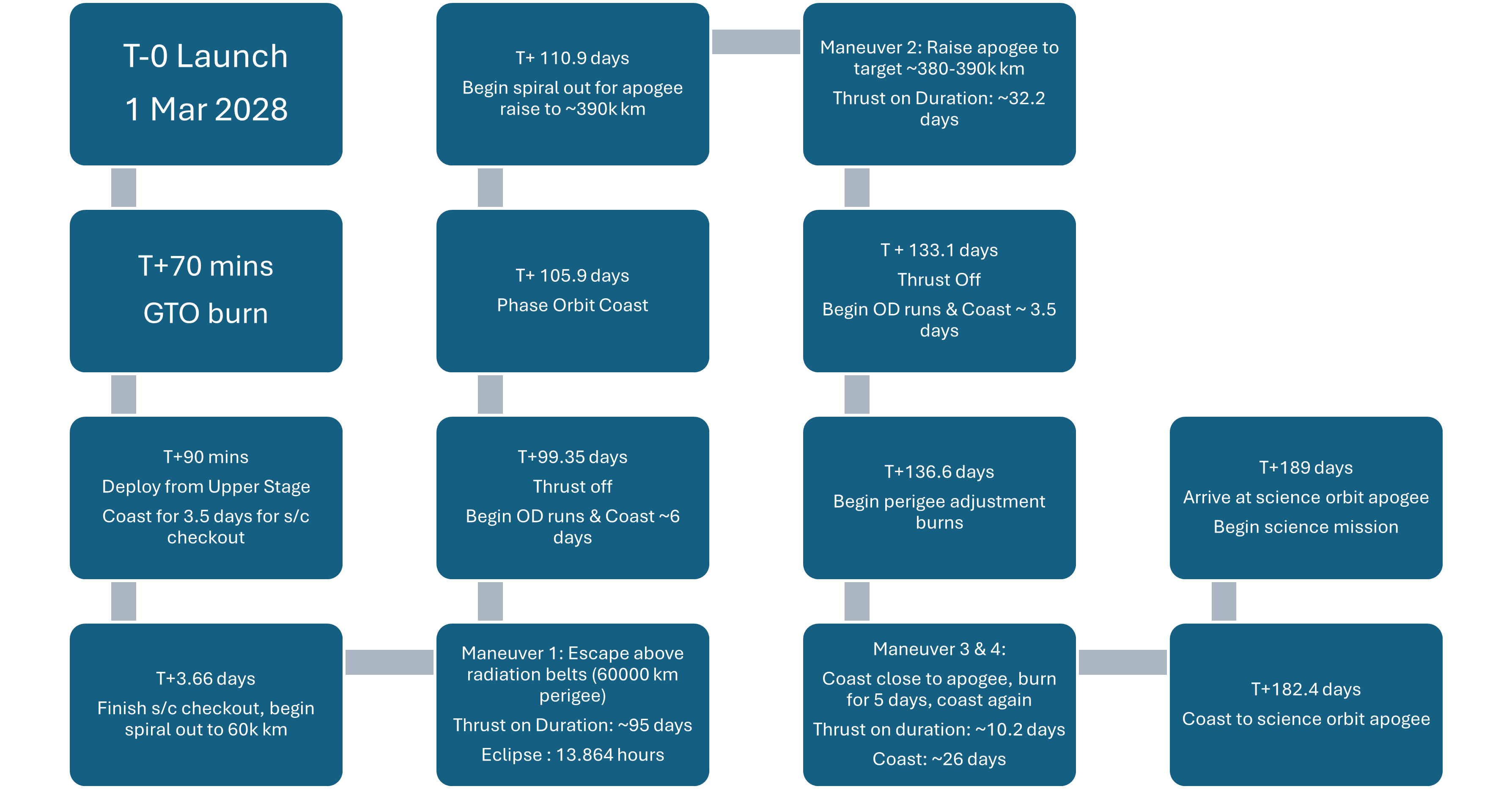}
    \caption{CONOPS for transfer to the 2:1 LRO from an SSTO}
    \label{fig:lro_conops}
\end{figure}

The converged trajectory for a transfer to a 2:1 LRO is shown in \cref{fig:lro_trajectory_converged}. We developed a nominal concept of operations (CONOPS) utilizing this transfer to the LRO as shown in \cref{fig:lro_conops} assuming a launch date of March 1st, 2028. Any changes to the launch date or initial mass configuration will change some of the duration of the burns and transfer times, but the general flow of events should stay the same. The spacecraft spends $\approx$ 100 days passing in and out of the radiation belts until the perigee is above 60,000 km. Our analysis shows that the final science orbit is stable for more than 50 years i.e. the perigee stays above 10$R_e$ (GEO is at 6$R_e$), mitigating any space traffic management concerns. The maximum eclipse period during the operational mission lifetime is 2.5 hours with a total of five eclipses in five years. 

\subsubsection{SEL2 Halo Results}
\begin{figure}[h!]
    \centering
    \includegraphics[width=0.95\linewidth]{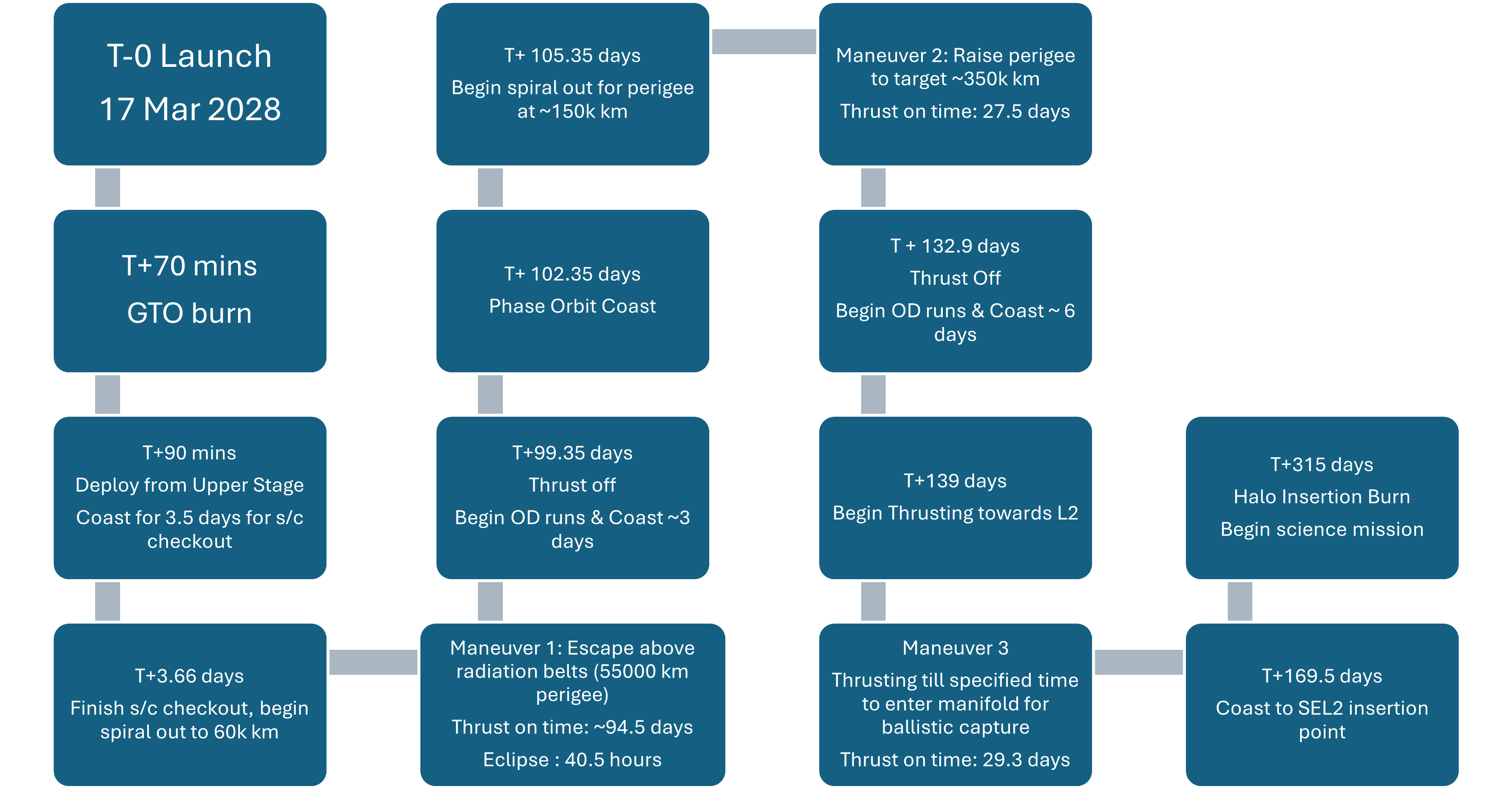}
    \caption{CONOPS for transfer to the \SELtwo{} Halo orbit}
    \label{fig:sel2_conops}
\end{figure}

The fully converged transfer orbit is shown in \cref{fig:sel2_conops}. This solution is for a given launch date and initial spacecraft parameters. Due to obtaining the transfer solution using a backward propagation method, the final state of the spacecraft at the entry point of the halo orbit was selected as the "initial" state including the value of fuel remaining at orbit insertion. If the halo orbit epoch or final spacecraft fuel mass values or engine parameters have changed, the transfer trajectory will look a bit different as well as some of the numerical results but generally should be similar to the values shown below.

A nominal CONOPS for the converged transfer trajectory is shown in \cref{fig:sel2_conops}. As the \textit{Astrogator} sequences were built backwards in time, the CONOPS is presented in a forward timeline as such the dates/times are an estimate. Any changes to the launch date or initial mass configuration will change some of the duration of the burns and transfer times but the general flow of events should stay the same. 
\begin{figure}[h!]
    \centering
    \includegraphics[width=0.99\linewidth]{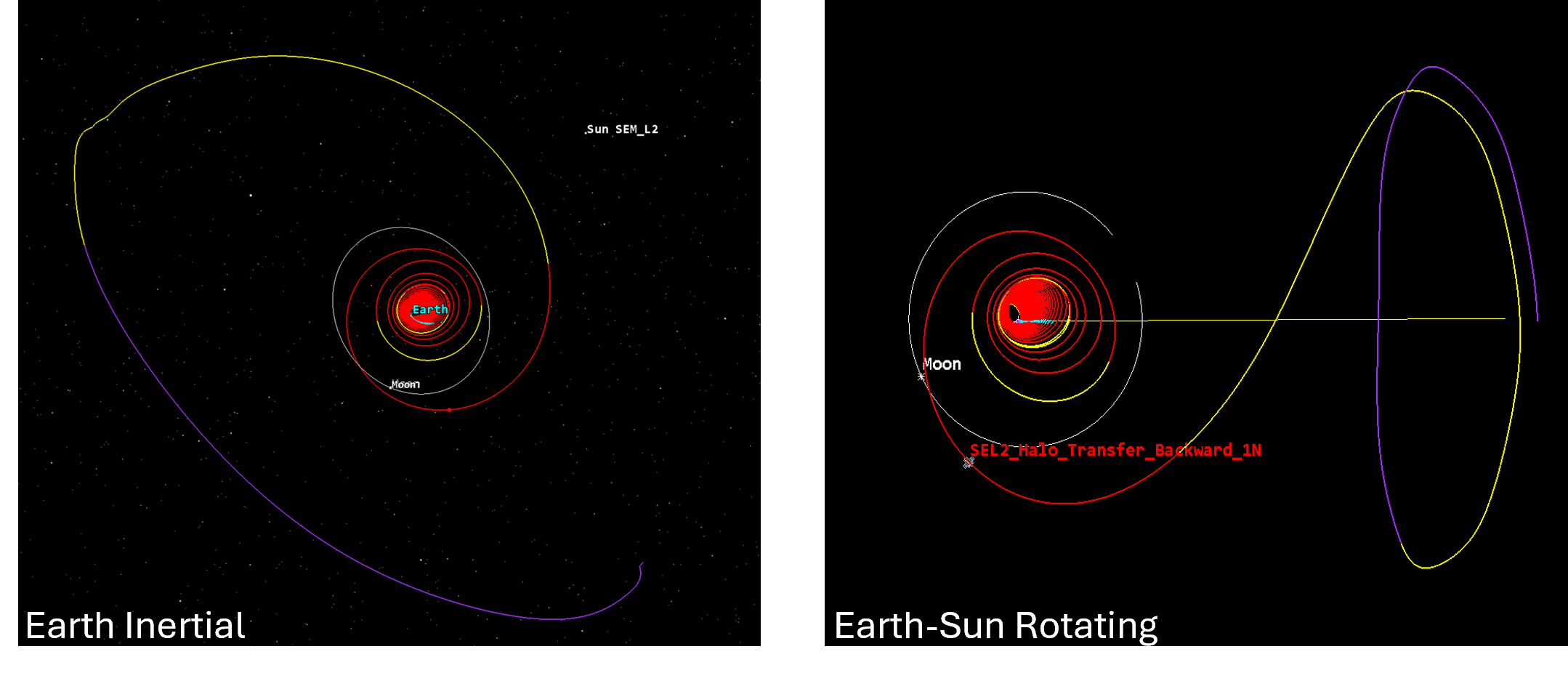}
    \caption{Converged transfer trajectory to a \SELtwo{} halo orbit in both Earth Inertial and Earth-Sun rotating reference frames. Thrusting segments are shown in red, and eclipse locations are shown in blue while other colors are coasting segments.}
    \label{fig:sel2_trajectory_converged}
\end{figure}

\subsection{Results of preliminary radiation study}

\subsubsection{LRO Radiation Results}
\begin{figure}
    \centering
    \includegraphics[width=0.99\linewidth]{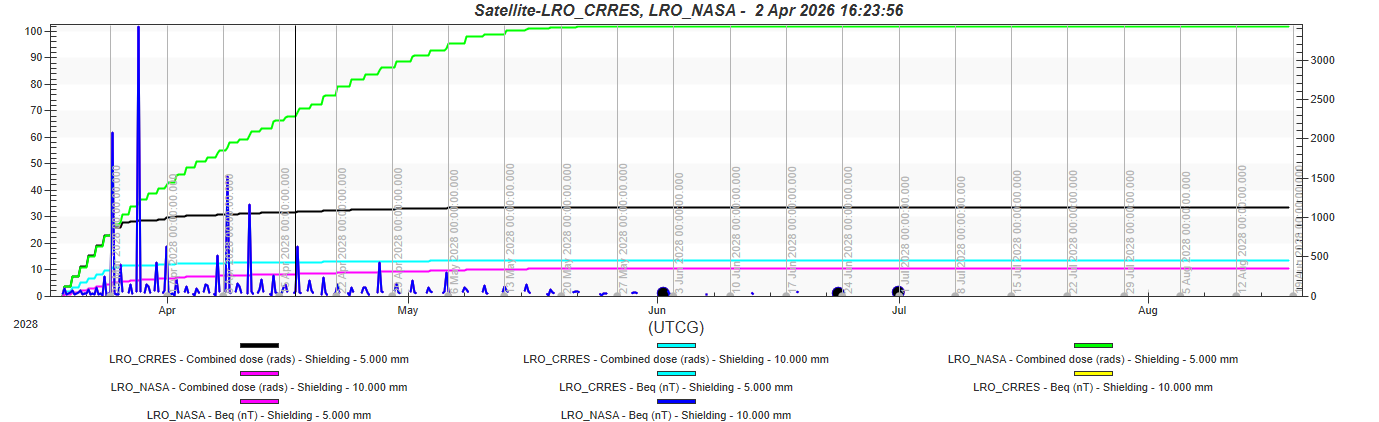}
    \caption{Combined dosage over time for LRO transfer}
    \label{fig:combined_dosage_lro}
\end{figure}

The total dose for the LRO transfer case is shown in \cref{tab:lro_radiation_results}. For the NASA computational mode, the LRO transfer trajectory was predicted to receive a total radiation dose of 158.6 rads by the time it exited the Van Allen radiation belts for the 5 mm shielding case, and 19.0 rads for the 10 mm case. For the 5 mm case, most of this radiation came from the electron dose, while the majority of total radiation in the 10 mm case consisted of the proton dose.
The CRRES computational mode predicted a narrower gap between the two shielding thicknesses, reporting a more conservative estimate of total dosage for the thicker shielding case and a less conservative estimate for the thinner shielding case. For both cases, CRRES predicted a smaller electron dose and higher proton dose than the NASA model. \cref{fig:combined_dosage_lro} shows the accumulated dosage over time.
\begin{table}[t!]
    \centering
    \caption{Total dosage for transfer to LRO using NASA and CRRES model}
    \begin{tabular}{|c|m{1.75cm}|m{1.75cm}|m{2cm}|m{1.75cm}|m{1.75cm}|}
    \hline
         & Shielding thickness (mm) & Electron dose (rads) & Electron-Bremsstrahlung dose (rads) & Proton dose (rads) & Combined dose (rads)\\
         \hline
        \multirow{2}{4em}{NASA model} & 5 & 111 & 10.22  & 37.36 & 158.6\\
        \cline{2-6}
         & 10 & 0.1828 & 6.386 & 12.44 & 19\\ \hline
         \multirow{2}{4em}{CRRES model}&5  &3.448  &3.540  & 65.69 & 72.68\\ \cline{2-6}
         & 10 & $9.242 \times 10^{-3}$  & 2.376 & 28.27 &30.65 \\ \hline
    \end{tabular}
    \label{tab:lro_radiation_results}
\end{table}

\subsubsection{\SELtwo{} Radiation Results}

The total dosage for the \SELtwo{} Halo orbit transfer case is shown in \cref{tab:sel2_radiation_results}. For the NASA computational mode, the SEL2 transfer trajectory was predicted to receive a total radiation dose of 151.7 rads by the time it exited the Van Allen radiation belts for the 5 mm shielding case, and 18.8 rads for the 10 mm case. As with the LRO transfer, most of the radiation for the 5 mm case came from electron dose while the 10 mm case experienced a larger proton dose.
The CRRES computational mode also predicted a narrower gap between the two shielding thicknesses for the \SELtwo{}  trajectory, reporting a combined dose of 73.8 rads for the 5 mm shielding case and 31.2 rads for the 10 mm case. For both trajectories, CRRES predicted a smaller electron dose and higher proton dose than the NASA model.

\begin{table}[h!]
    \centering
    \caption{Total dosage for transfer to \SELtwo{}  halo orbit using NASA and CRRES model}
    \begin{tabular}{|c|m{1.75cm}|m{1.75cm}|m{2cm}|m{1.75cm}|m{1.75cm}|}
    \hline
         & Shielding thickness (mm) & Electron dose (rads) & Electron-Bremsstrahlung dose (rads) & Proton dose (rads) & Combined dose (rads)\\
         \hline
        \multirow{2}{4em}{NASA model} & 5 & 104.9 & 9.863  & 36.94 & 151.7\\
        \cline{2-6}
         & 10 & 0.1674 & 6.169 & 12.48 & 18.81\\ \hline
         \multirow{2}{4em}{CRRES model}&5  &3.071  &3.527  & 67.22 & 73.82\\ \cline{2-6}
         & 10 & $7.623 \times 10^{-3}$  & 2.368 & 28.80 &31.17 \\ \hline
    \end{tabular}
    \label{tab:sel2_radiation_results}
\end{table}

For both trajectories, it is worth noting that while the NASA model sees radiation continuously accumulate until the satellite has fully exited the Van Allen radiation belts for the 5 mm shielding case while the 10 mm NASA case and both CRRES cases find the rate of radiation accumulation flattens as perigee exceeds 5000 km,  roughly the upper limit of the lower Van Allen belt. 

\begin{figure}
    \centering
    \includegraphics[width=0.99\linewidth]{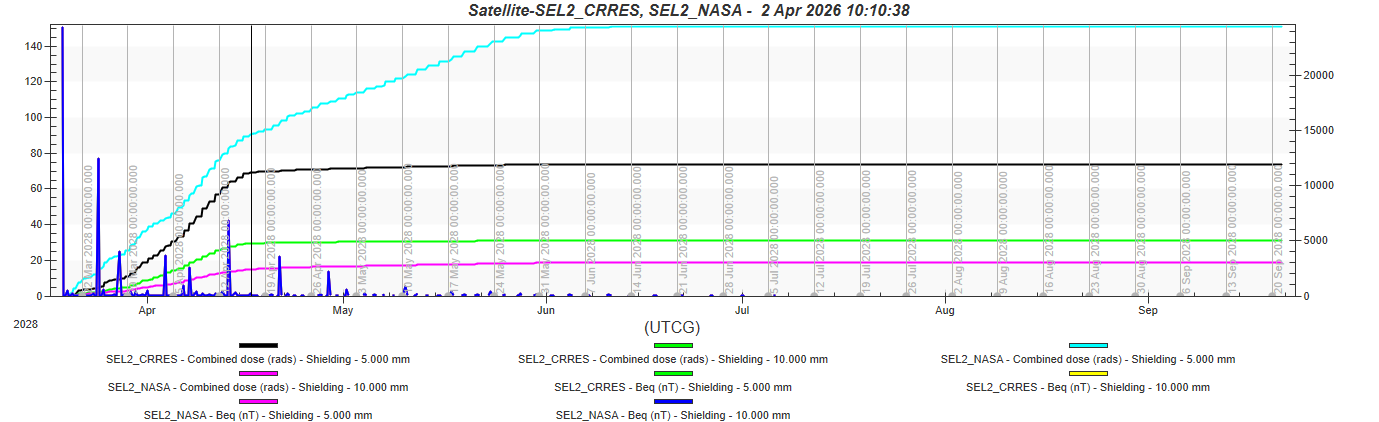}
    \caption{Combined Dosage over time for transfer to \SELtwo{}  Halo orbit}
    \label{fig:combined_dose_sel2}
\end{figure}

\begin{figure}
    \centering
    \includegraphics[width=0.95\linewidth]{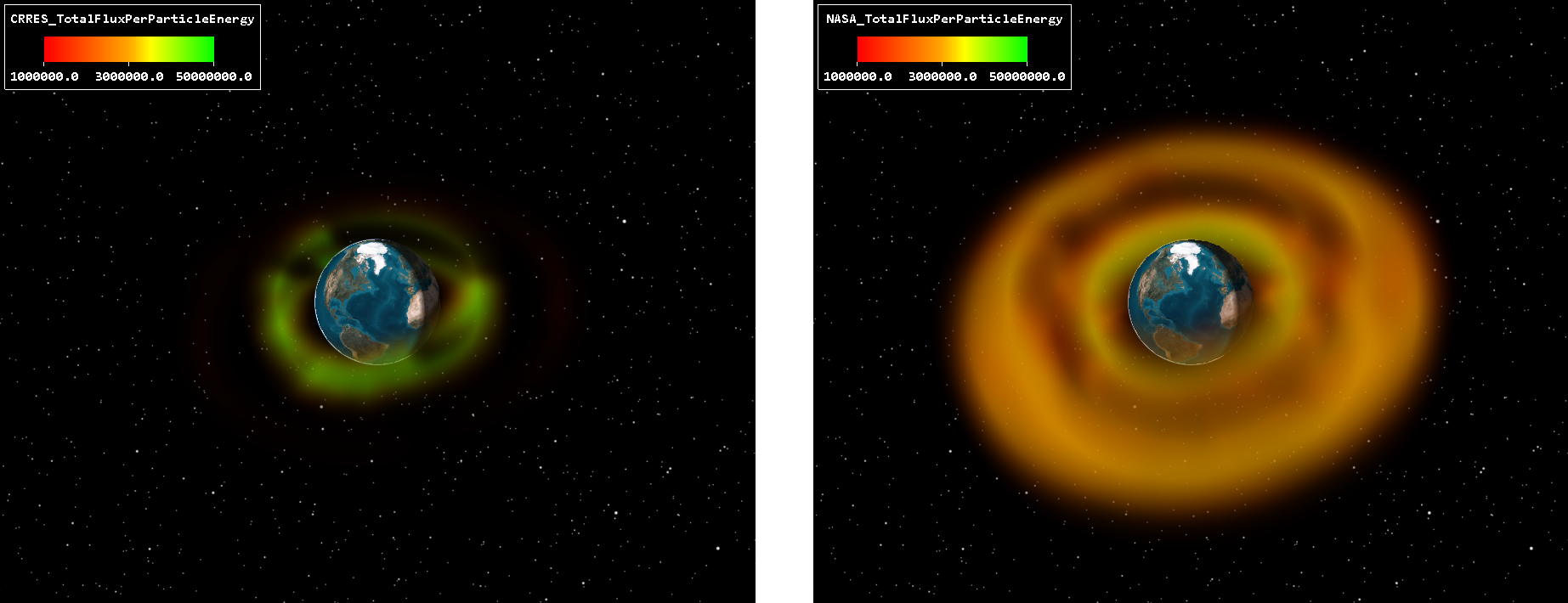}
    \caption{Visualization of Total Flux for CRRES model (left) and NASA model(right)}
    \label{fig:totalfluxmodel}
\end{figure}

Assuming for the lowest shielding case investigated (5 mm Al), the minimum energy of a proton to pierce this shielding is roughly 30-35 Mev \cite{burrell_calculation_1965}, and the minimum electron energy is roughly 5 MeV \cite{lazurik_average_1998}, we find that although the upper belt exhibits high radiation flux, most of this radiation has insufficient energy to pierce the minimum shielding case.
We can visualize this in STK by defining two satellites with the CRRES/NASA radiation models respectively shown in \cref{fig:totalfluxmodel}. We then extract a calculation element for the flux per particle energy for each particle energy value that is above the lower limits for penetration of 5 mm shielding. We then add these values together to roughly approximate the “total flux” of particles with sufficient energy to penetrate the defined shielding on a volumetric grid between the two models: 
We thus see that although both models find the most intense radiation environment for the given shielding in the lower Van Allen belt, the NASA model tends to predict a larger proportion of high-energy particles– primarily high energy electrons– in the upper Van Allen belts than the CRRES model. Furthermore, since 10 mm shielding is sufficient to negate even the highest energy electrons modeled by either the NASA or CRRES SEET modes– up to roughly 10 MeV– we find that all 10 mm shielding cases receive the majority of their radiation dose from higher-energy protons in the lower Van Allen belt \cite{lazurik_average_1998}.

\subsection{Results of link budget analysis}
\begin{figure}[t!]
    \centering
    \includegraphics[width=0.85\linewidth]{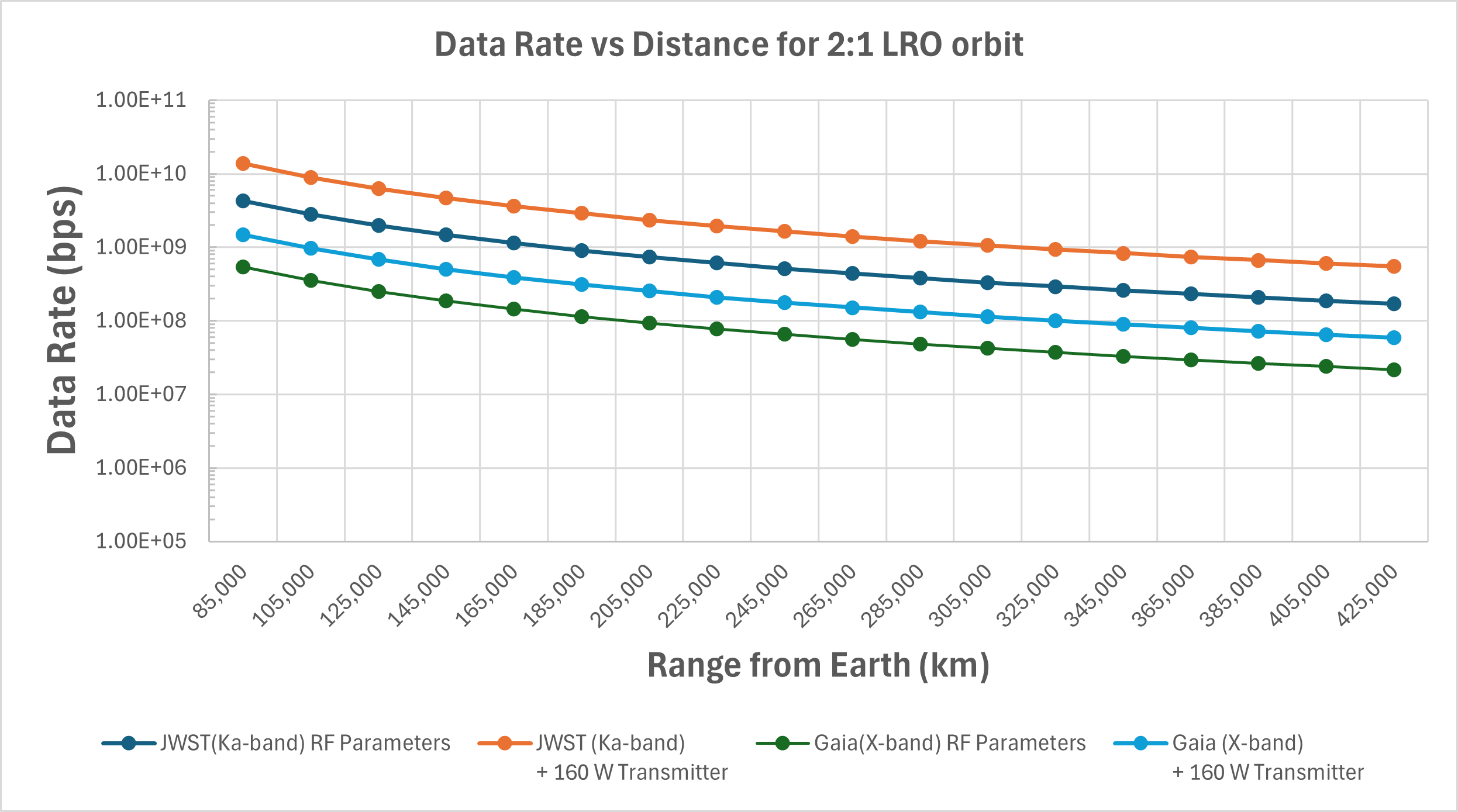}
    \caption{Data rates for Lunar Resonant Orbit vs distance from Earth}
    \label{fig:lro_datarate}
\end{figure}

\begin{figure}[h!]
    \centering
    \includegraphics[width=0.85\linewidth]{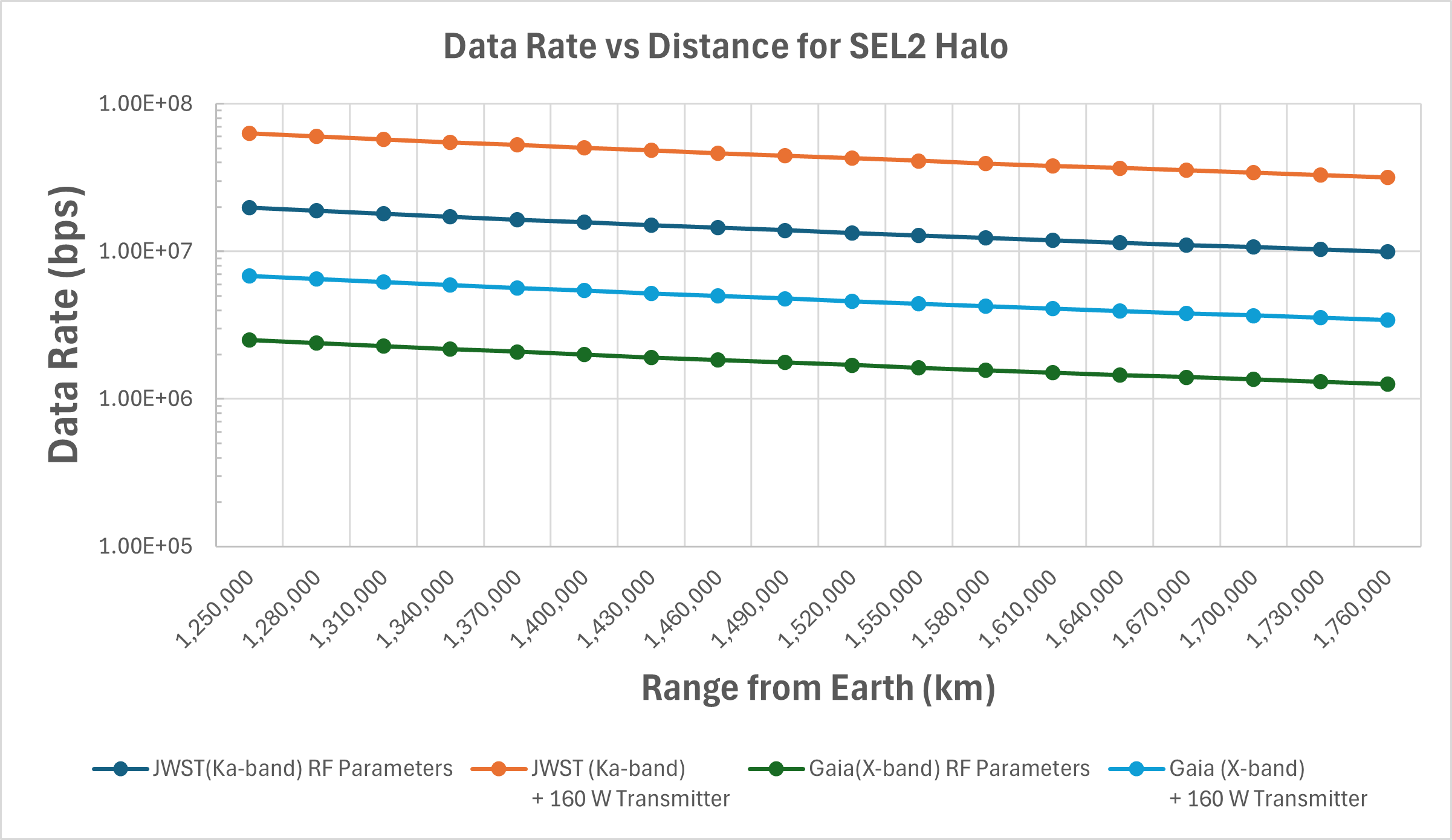}
    \caption{Data rates for Lunar Resonant Orbit vs distance from Earth}
    \label{fig:sel2_datarate}
\end{figure}

Our maximum supportable data rates in the LRO and in \SELtwo{} are shown in \cref{fig:lro_datarate,fig:sel2_datarate}. In both cases, the Ka-band system offers higher data rates. However, the Ka-band system on JWST utilized a dish antenna compared to the phased array antenna on GAIA. The dish provided a higher gain than the phased array which also contributes to the increased data rates that we see in the Ka-band case. As mentioned earlier, our team has not baselined the type of spacecraft antenna to be used and as such the analysis must be updated with RF parameters for our spacecraft. While Ka-band offers higher data rates, there are many more deep space X-band ground stations available with flight heritage compared with Ka-band. As such, a more refined trade must be performed by discussing with ground station providers on future capabilities. Using conservative values for data rates from the results shown, we expect the following minimum averaged daily data downlink volume for LRO of 3.46 Tb (X-band)/28.9 Tb (Ka-band) and \SELtwo{} halo orbit case, 0.3 Tb (X-band)/0.86 (Ka-band). From this analysis it is clear that the LRO location offers much greater payload data volume downlinked compared with the \SELtwo{} halo orbit case even when using the slower data rate option for the LRO. 


\subsection{Results of science yield comparison}

\begin{figure}[H]
    \centering
    \begin{minipage}{0.48\textwidth}
        \centering
        \includegraphics[width=\linewidth]{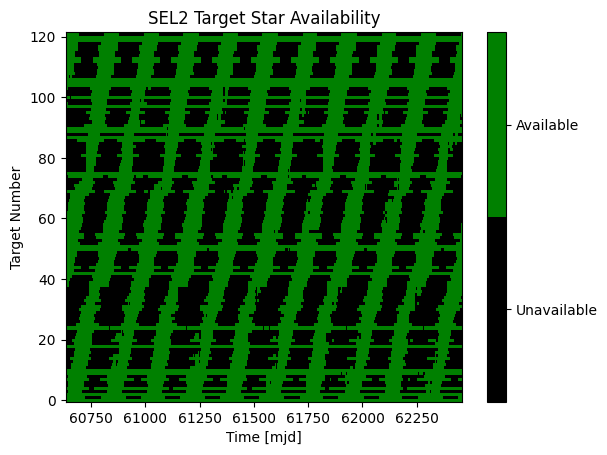}
    \end{minipage}
    \hfill
    \begin{minipage}{0.48\textwidth}
        \centering
        \includegraphics[width=\linewidth]{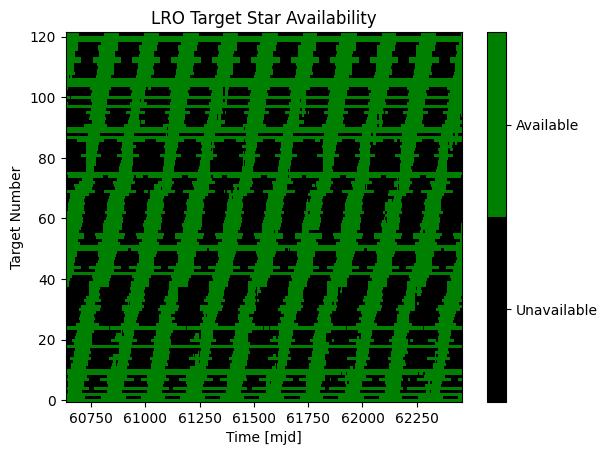}
    \end{minipage}
    \caption{Keepout maps for \SELtwo{} (left) and LRO (right), generated from availability of target stars due to keepout angle constraints. These keepout maps are used for ensuring efficient observation scheduling and understanding keepout restrictions from each orbit.}
    \label{fig:keepoutMaps}
\end{figure}

Running the EXOSIMS simulation builds a keepout map that is determined from the keepout angle constraints specified. Figure~\ref{fig:keepoutMaps} displays our keepout maps generated for both the \SELtwo{} orbit and the LRO. These maps show the times that targets are available/unavailable throughout the mission lifetime, with each row representing one star. This availability is used to determine when visits to each target can be scheduled for survey efficiency and allows us to estimate expected yield. The targets with longer availability windows are easier to revisit for maximizing observational completeness.

\begin{figure}[b!]
    \centering
    \begin{minipage}{0.48\textwidth}
        \centering
        \includegraphics[width=\linewidth]{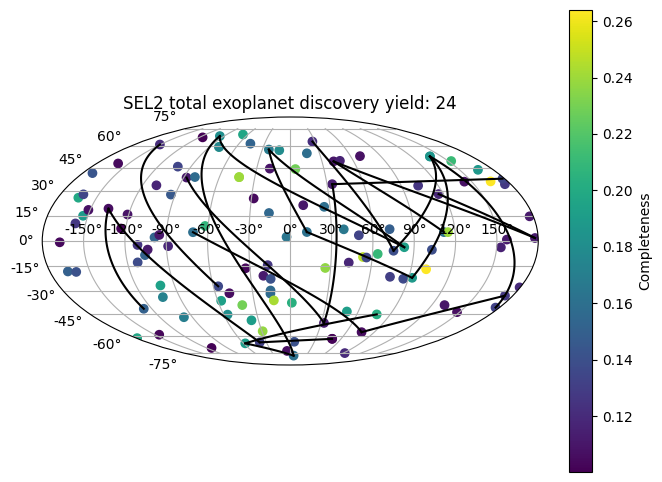}
    \end{minipage}
    \hfill
    \begin{minipage}{0.48\textwidth}
        \centering
        \includegraphics[width=\linewidth]{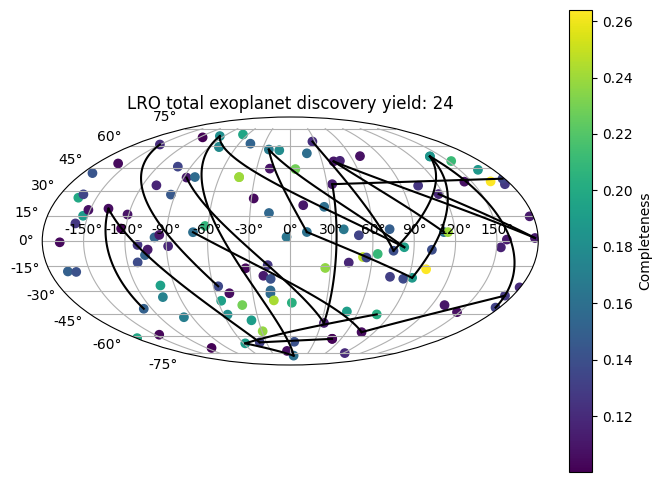}
    \end{minipage}
    \caption{Completeness maps for \SELtwo{} (left) and LRO (right). These maps demonstrate the locations in the sky where target stars have the most completeness and the black lines connecting them show the sequence of observed targets. This determines the areas in the sky where targets are most observable allowing for optimal observation sequencing for maximizing completeness.}
    \label{fig:completenessMaps}
\end{figure}
Figure~\ref{fig:completenessMaps} demonstrates the completeness maps generated by simulations for each orbit. The map is built from a Mollewide projection with the right ascension and declination plotted for each target star. The line segments connecting the stars indicates the sequence of observed targets and the colors indicate the completeness. The map demonstrates where in the sky the high-completeness targets are and how the scheduler moved among them. It demonstrates that the mission spends the most time in the most accessible regions, where the targets lie within the field of regard. Moreover, it demonstrates the tradeoff between maximizing completeness and maximizing practicality in observation sequencing. This map is as expected; the areas where stars are most accessible have the highest likelihood of exoplanet detection \cite{completenessSPIE}. 
\begin{wrapfigure}[29]{r}{.5\textwidth}
   \centering
    \includegraphics[width=0.95\linewidth]{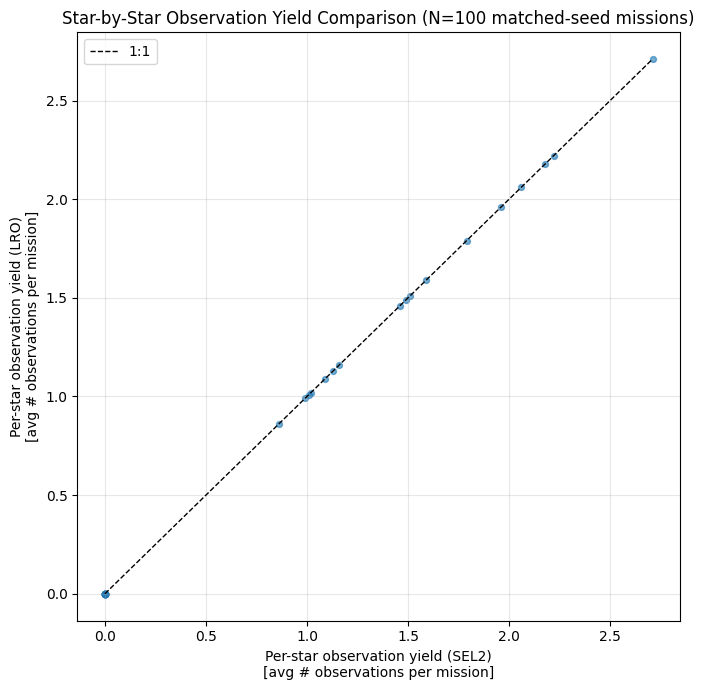}
    \caption{Comparison of average number of observations per mission for 100 missions ran. The data points shown are the number of successful observations for each mission, with the x-axis plotting the yield for the \SELtwo{} halo orbit and the y-axis plotting the yield for the LRO. The dashed line shows where observations were successful for both orbit architectures to make a comparison of science yield.}
    \label{fig:yieldLROandSEL2}
\end{wrapfigure} 
A comparison between the number of exoplanets discovered is shown in \cref{fig:yieldLROandSEL2}. This is a star-by-star scatter comparison of the observation yield between the \SELtwo{} halo orbit (horizontal axis) and LRO (vertical axis), with the dashed line representing an equal number of exoplanets found. It was generated by running 100 missions for each orbit configuration and counting how many stars were successfully observed, then averaging the values for each mission. It demonstrates the stars that were favored by one orbit architecture over the other.

We determined that the keepout angles and field of regard have more of an impact on exoplanet yield than the orbit geometry. Due to the extreme similarities between the quantity of exoplanets detected from \SELtwo{} and from LRO, we can confirm that the yield between the two orbits was fairly similar. When changing some of the specifications of our telescope, such as increasing the diameter and contrast, we do see small differences between the yields of each orbit architecture. Future work will involve quantifying these exact differences and determining which combination of factors causes the widest relative difference. Thus far, we have found that both orbit architectures result in similarly successful missions to detect exoplanets. Due to the similar output in the yields between the two, selecting a closer orbit such as LRO may be  cheaper and would allow for more payload mass. Our results show that a larger telescope diameter always results in higher yield of exoplanets. Having the freedom to place a large telescope into a more accessible orbit relative to \SELtwo{} would be incredibly valuable. Because the LRO yield matches that of the \SELtwo{} halo orbit yield, mission design may not need not be limited to a Lagrange point, as the LRO is closer and cheaper to reach, while producing similar science return.

\section{Conclusion}
This work investigates how orbit selection influences the exoplanet detection yield of a 3-meter class space telescope. We determined that both orbit choices result in similar exoplanet yields due to keep out angles and field of regard dominating the output. Due to the advantageous thermal stability and orbit architecture of both options, we can conclude that both prove operational for exoplanet detection. This allows us to prioritize mission cost when it comes to deciding between the two orbits, and allows flexibility for transfer time and payload mass. We also show that with low thrust systems, transfers to orbits like the LRO and \SELtwo{} halo are achievable with current spacecraft technology. In future work, we plan to broaden the mission specifications further to determine what parameters specifically impact the difference in scientific yield between the \SELtwo{} halo orbit and the LRO. In addition we plan to include thrust steering laws to optimize the thrust direction calculation and further decrease fuel usage during transfers. We also would like to examine the transfer trajectory options for different launch dates across the year.

\section*{Acknowledgments}
Portions of this research were supported by funding from the Technology Research Initiative Fund (TRIF) of the Arizona Board of Regents and by the Steward Observatory of University of Arizona.

\newpage
\bibliography{references,exoplanets}

\typeout{get arXiv to do 4 passes: Label(s) may have changed. Rerun}

\end{document}